1# Bundling Impacts in Spanish Telecommunications Market
Juan J. Marfil, Antonio Sanchez, Belen Carro. U. Valladolid
Using abstract tag.Abstract— Bundling is a pervasive strategy in telecommunications markets. Due to the great success obtained at Spanish Telco market, this article analyzes the impacts of the introduction of convergent bundling on that market. This commercial strategy has caused major changes in the market, both from the point of view of demand and its own structure. From the demand-side, we analyze how product bundling has produced a substantial reduction on effective prices in the sector, and by means of technological diffusion models, we show that bundling is the main factor of Fixed Broadband development, along with FTTH networks deployments, in contrast with stand-alone product marketing scenarios.  Pay TV inclusion and hard-bundling strategies at communications packages has motivated to the greatest growth of this product in the history of this market. Regarding market structure, the introduction of convergent bundling has led to a consolidation in the market, increasing the concentration around the main convergent players
1# Bundling Impacts in Spanish Telecommunications Market

Juan J. Marfil, Antonio Sanchez, Belen Carro. U. ValladolidAbstract— Bundling is a pervasive strategy in telecommunications markets. Due to the great success obtained at Spanish Telco market, this article analyzes the impacts of the introduction of convergent bundling on that market. This commercial strategy has caused major changes in the market, both from the point of view of demand and its own structure. From the demand-side, we analyze how product bundling has produced a substantial reduction on effective prices in the sector, and by means of technological diffusion models, we show that bundling is the main factor of Fixed Broadband development, along with FTTH networks deployments, in contrast with stand-alone product marketing scenarios.  Pay TV inclusion and hard-bundling strategies at communications packages has motivated to the greatest growth of this product in the history of this market. Regarding market structure, the introduction of convergent bundling has led to a consolidation in the market, increasing the concentration around the main convergent players
**Index Terms**—Bundling, Convergence, broadband diffusion, pay TV, Spanish Market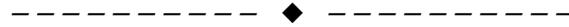

## 1 INTRODUCTION

In recent years, the practice of bundling has been introduced within the telecommunications industry. At European level, 50% of EU households purchased packaged telecommunications services in 2015 (European Commission, 2015). This trend has had a much greater impact on the Spanish market. At the same year, more than 70% of households in Spain purchased triple-play or quad-play telecommunications and payTV bundles (CNMC, annual report 2016).

The objective of this study is firstly to provide a contextual framework that explains the origin of bundling/convergence in the Spanish market and the reasons why it has presented such level of adoption and subsequently, analyze the impacts it has produced over the market. The impacts have been proposed as hypotheses and confirmed by analytical studies. The first effect of the convergence analyzed is its influence on pricing of communications services and whether bundling has meant a reduction in the effective price of the sector's products. Second, we analyze whether bundling has increased the adoption of Broadband services, using different diffusion models, as well as the growth in pay TV penetration thanks to the introduction in a "hard-bundle" mode inside of telecommunications services. Finally, we analyze the impact of bundling within the market structure (as driver in the consolidation of companies in the sector).

The rest of the paper is structured as follows.  Section 2 provides a literature review on bundling. Section 3 presents an overview of the Spanish telco market, describing the main agents and its competitive dynamics prior to the massive launch of convergent quad-play bundle. Section 4 introduces the main hypotheses and their validation regarding the impact of convergence and bundling on this market, both in terms of demand (evolution of pricing, fixed broadband diffusion and growth in pay TV) and market structure (consolidation and competition). Finally, section 5 provides the main conclusions of this study.

## 2 LITERATURE REVIEW

Bundling – initially defined as two or more products (goods or services) that are available for purchase as one single product (Guiltinan, 1987) - is a widespread commercial practice in many industries, from retail to banking sector.
In the last few years, this strategy has also been applied to in the telecommunications industry and was developed in several stages.  Initially, bundling was made up of different communication services in order to encourage product adoption (e.g., internet broadband services + telephone, mobile voice + sms / mobile broadband, pay TV and telephony) as a type of loyalty programs (e.g., mobile communications + mobile phone subsidies) or as a lever to customer develop beyond communications services (mobile services that include added value services like music, mobile games, etc).  This bundling strategy in media and telecommunications has witnessed a more recent change with the development of a quad-play offer, a bundle that combines the triple play service of internet broadband service, pay TV and telephone with mobile communications.  That offer emerged as the reference offer in many markets, especially in the Spanish one, as will be analyzed in detail throughout the article

"© 20xx IEEE. Personal use of this material is permitted. Permission from IEEE must be obtained for all other uses, in any current or future media, including reprinting/republishing this material for advertising or promotional purposes, creating new collective works, for resale or redistribution to servers or lists, or reuse of any copyrighted component of this work in other works."

In the following lines, the different types of bundling using the categories most commonly employed in the literature (Stremercsh, S & Tellis, G., 2002), are described; and, secondly, the types of bundling related to the Spanish telco market are presented.

Bundling is defined as the sale of two or more separate products in one package. Separate products are understood as products for which separate markets exist, because at least some buyers are willing to buy or actually buy the products separately. For example, in mobile services, mobile phone can be purchased separately from mobile communications services (commonly called sim-only services); in the field of internet broadband services, it can be distinguished between "naked" broadband, where only the internet broadband connection is available, and that which includes telephony services

The different categories used are reviewed throughout the article:
- Price bundling: sale of two or more separate products in a package at a discount, without any integration of the products.
- Product bundling: integration and sale of two or more separate products.

The distinction between both categories is particulary relevant since it entails different strategies and considerations for the companies that undertake them. The Price bundling strategy is more tactical, requires less operational integration (systems, marketing channels, etc.) and is a clear promotional tool, so the perceived customer value may be lower, while the Product bundling strategy is long-term and differential, but it requires a greater operational complexity (integration of products and systems in many levels: customer care, sales, fulfillment, invoicing, delivery, etc.) although the perceived customer value is higher.

Bundling form (Adams, W. & Yellen, J., 1976) is another concept considered in order to define the bundling firm´s strategy:
- Pure bundling: a firm sells only the bundle and not (all) the products separately.
- Mixed bundling: a firm sells both the bundle and (all) the products separately.
    - Unbundling: as opposed to previous options, a firm only sells the products separately.

In addition, there is a point to be considered and not previously addressed in these models, but that has been decisive in the Spanish market evolution: how the bundling is commercialized.
    - Opt-in Bundling: when customers add functionalities or capabilities through add-ons to their commercial offer to configure the complete offer.
    - Hard Bundling: when the firm includes in the bundle all the main components of the product, without any alternatives in other configurations for the same characteristics or benefits.

Although the main objective of bundling is to increase revenue by means of extracting more surplus from heterogeneous consumers and demand, there are other specific benefits associated with telecommunications sector that have also been analyzed, such as the capture of new clients from competitors, increase customer base loyalty and reduce the churn (Prince & Greenstein, 2014; Yang, 2013) or leverage operators position in the market through product tying.

Regarding how demand is affected, recently Sobolewski, M. & Kopczewski, T. (2017) have estimated willingness to pay for the three major fixed-line telecommunication services (telephony, broadband and payTV) offered as stand-alone services or in different dual o triple-play bundles, identifying that broadband is the biggest value generating service in fixed network and offers a potential for additional value creation from integration with payTV.

Srinuan, Srinuan, & Bohlin, 2014 have analyzed what are the determining factors that explain the probability that a consumer buys multiple services, identifying product discount and a high level of customer incomes as main factors for consumer buys a bundling service.

In the Spanish telco market, the communications offer was initially based on a price and mixed bundling model with cross discounts as new communications products were included in the communications package (mainly fixed broadband and mobile and / or TV), until the launch of a quad-play bundle by the main telecommunications operator (Telefonica) in 4Q2012 which caused a massive adoption of this type of bundling by the rest of players. As will be seen in section 4.1, the commercialization in hard-bundle mode will serve as a fundamental lever to explain the growth and development of pay TV in recent years, when it was introduced within these quad-play communications offer.

## 3 SPANISH TELECOMMUNICATIONS MARKET

Spanish telecommunications market was traditionally based on a large number of players offering telephony,

broadband internet services, wireless communications and pay TV. This competitive environment, dependent on the different agent capacities, was composed of large players (multinational groups), smaller agents with national and/or local character focused on market niches and agents specialized in the payTV business.

Table 1 summarizes the agents and their main characteristics in order to understand the competitive landscape, as well as their networks capacities, a key issue to comprehend future market evolution.

TABLE 1
SPANISH MARKET TELCO PLAYERS

| Company | Main Offer | Network Capabilities |
| --- | --- | --- |
| Telefonica | FBB, Mobile, TV | 100% Footprint (FBB & Mobile) |
| Vodafone | FBB, Mobile | 100% Footprint Mobile; ULL |
| ORANGE | FBB, Mobile | 100% Footprint Mobile; ULL |
| Yoigo | Mobile | Mobile Footprint & MVNO |
| ONO | FBB, Mobile, TV | MVNO; HFC |
| Jazztel | FBB, Mobile | MVNO; ULL |
| Euskaltel | FBB, Mobile, TV | MVNO; HFC regional footprint |
| Telecable | FBB, Mobile, TV | MVNO; HFC regional footprint |
| R | FBB, Mobile, TV | MVNO; HFC regional footprint |
| MVNOs | Mobile | MVNO |
| GolTV | TV | Pay TV - TDT |
| Digital + | TV | DTH 100% Footprint |

Telco offer was based mainly on stand-alone communications packages: Internet Broadband services, Wireless communications, telephony, pay TV, with some bundling between them. In some cases, local cable operators, Euskaltel, R, and Telecable, along with the main national cable company, ONO, offered quad-play bundles where all services were packaged, but the main activity of the telecommunications market was based on Internet Broadband services and Wireless communications. Table 2 summarizes the market shares by type of services for the different agents in 2011, prior to the massive launch of convergent offers at the end of 2012. The source of these data as well as the majority used throughout this article is the Spanish National Regulatory Authority (CNMC)[1].

TABLE 2
TELCO & PAY TV MARKET SHARES (2011)

| Company | Internet BB Services | Wireless Comms | Pay-TV |
| --- | --- | --- | --- |
| Telefonica | 49% | 39% | 18% |
| Vodafone | 8% | 28% | |
| ORANGE | 11% | 21% | 2% |
| Yoigo | - | 5% | - |
| ONO | 14% | | 21% |
| Jazztel | 10% | | - |
| Euskaltel | 2% | 7% | 4% |
| Telecable | 1% | | 3% |
| R | 2% | | 2% |
| MVNOs | - | | |
| Digital + | - | - | 39% |
| GolTV | - | - | 8% |

During 2012, there were three relevant events that would motivate a change in the sector and the Spanish Telco market, without which future evolution cannot be understood. These movements were initiated and led by the main market operator, Movistar (Telefonica Group). The rest of competitors followed it in their commercial strategy, to a greater or lesser extent, depending on their capabilities and /or commercial strategy. Firstly, the elimination or reduction of mobile handset subsidies. The commercial dynamics of the Spanish mobile market were based on subsidies of mobile handsets by the operators, allowing retaining or winning customers right before their contracts expired, renewing the terminal in exchange for a permanence commitment. The elimination of subsidies allowed the reduction of that commercial cost, and in some cases, the transference of these lower commercial costs to a price reduction for customers. Secondly, Movistar started a firm commitment to deploy a Fiber To The Home network (FTTH). By 2012, the country's FTTH footprint was 1.5 million households, in which New Generation Networks (NGN) main coverage came from cable operators (9 million households with Hybrid Fiber Coaxial -HFC- technology in 2011). This strong investment in FTTH network deployment, together with the quad-play bundling including pay TV as reference offer, changed the competitive landscape of the sector, leading to great changes as will be analyzed throughout the article.

## 4 BUNDLING IMPACTS

This section studies the different impacts and effects of the bundling (convergence communications package) on the Spanish Telco market. It will analyze the effects on the demand side, by means of the impact on pricing evolution, dynamization and diffusion of broadband technology as well as its impact on pay TV market by performing a hard-bundling commercial strategy as a general practice of marketing. Additionally, it will analyze the impact of bundling on market structure and competence. To date, there have been several previous studies analyzing the impacts of bundling/convergence on its different facets in a specific Telco market, like the Australian market where Papandrea, Stoceckl and Daly (2003) analyze the social welfare effects of bundling in that market, and the main factors that cause it (demand, marginal cost, economic distribution, pricing menus, market structure and fixed costs), the Turkish market (Üner, M., Güven, F & Cavusgil, T, 2015) and the Swedish market (Srinuan et al., 2014).

### 4.1 Impacts on demand evolution

In this section, we formulate three hypotheses and their corresponding validation about the impact of bundling on the demand side.

*Hypothesis 1: The introduction of bundling (convergent offers) in the market increased the effective price reduction at the Spanish telecommunications industry.*

Stigler (1963) was the first economist to recognize that a seller may be able to exercise price discrimination and increase benefits by selling product bundling rather independent products. According to Schmalensee (1984), bundling allows companies to offer a greater value proposition to the customer and exploit price discrimination, obtaining a higher consumer surplus than simple products, both in the pure bundling strategy and mixed Bundling. In the case of mixed bundling, it combines the advantages of reducing effective heterogeneity among those buyers with a high reserve price[2] for bundle products, while allowing the sale with a high mark-up for those buyers willing to pay a higher price for a single product.

In order to assess the impact on price due to bundling, we have analyzed the evolution of Average Revenue Per User (ARPU) of the products that make up the package: Telephony or Fixed Access, Fix Broad Band for each technology, xDSL, HFC and FTTH, wireless communications in both prepay and postpaid models, and pay TV, also in both technologies, based on IPTV or satellite (Direct to the Home - DTH) on a quarterly basis since 2009. This sector presents a clear deflationary trend, with a tendency to drop prices, year by year. In order to find out whether bundling has caused an additional ARPU reduction, we have built up the equivalent ARPU for the five most characteristic bundles on the market
- Bundle 1: telephony, Wireless communications (postpaid) and xDSL Broad Band.
- Bundle 2: telephony, Wireless communications (postpaid) and HFC Broad Band.
- Bundle 3: telephony, Wireless communications (postpaid), xDSL Broad Band and IPTV pay TV.
- Bundle 4: telephony, Wireless communications (postpaid), HFC Broad Band and pay TV.
- Bundle 5: telephony, Wireless communications (postpaid), FTTH Broad Band and IPTV pay TV.

Quarterly evolution of the equivalent ARPUs per bundle is illustrated in Figure 1.
Table 4 shows the year-over-year ARPUs variation for each bundle, showing a higher drop level during the years 2013 vs 2012 and 2014 vs 2013, concurring with the convergent bundle launching in 4Q12.

---

(2) The reservation price of a product is the maximum price a consumer is willing to pay for the product

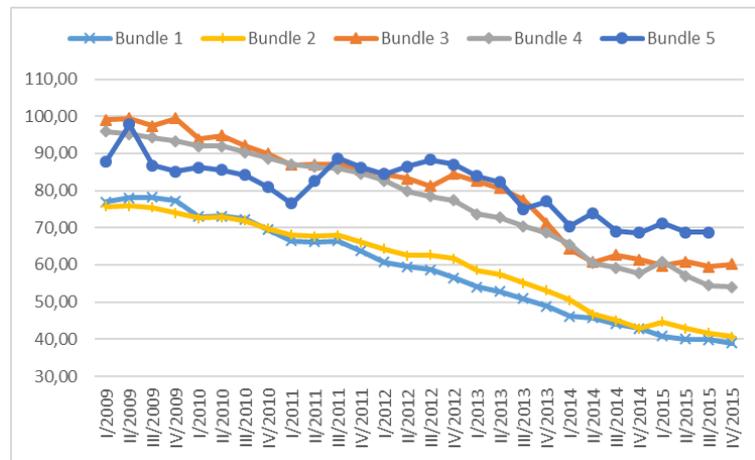

Fig. 1. Average Revenue per User (€) evolution per bundle type

TABLE 4
ARPU VARIATION (YEAR-OVER-YEAR) PER BUNDLE

|  | % yoy 08-09 | % yoy 09-10 | % yoy 10-11 | % yoy 11-12 | % yoy 12-13 | %yoy 13-14 | %yoy 14-15 |
|---|---|---|---|---|---|---|---|
| Bundling 1 | -2% | -10% | -8% | -11% | -14% | -12% | -9% |
| Bundling 2 | -4% | -10% | -5% | -1% | -16% | -14% | -2% |
| Bundling 3 | -7% | -5% | -5% | -8% | -11% | -16% | -6% |
| Bundling 4 | -7% | -6% | -5% | -7% | -14% | -19% | -5% |
| Bundling 5 | - | -3% | 5% | -1% | -15% | -8% | 0% |

In order to discriminate the increase of total ARPU drop caused by launching convergent packets, we have analyzed the evolution of the different ARPUs in this time series, using Ordinary Least Squares correlation method. Initially, the analysis was divided into two periods:
- 1Q2009 to 3Q2012: pre-convergence period, where the commercialization of stand-alone products was predominant.
- 4Q2012-4Q2015: convergence period, where quad-play bundles are widely offered and monopolize virtually all of the commercial offer.

The results are summarized in Table 5, where bundles 1, 2, 3 and 4 have high correlation levels (R2> 0.94 in bundles 1, 2 and 4, and R2> 0.83 in bundle 3) and clearly pass the Fisher test with a significance level α = 0.05. For the four cases, the slope value obtained (parameter m) undergoes a decrease in the convergence period versus the pre-convergence period, validating the hypothesis and confirming a higher ARPU fall during this period versus the previous trend.

TABLE 5
RESULTS OF REGRESSION PER TYPE OF BUNDLING

|  | 1Q09-3Q12 | 4Q12-1Q15 | 4Q12-4Q15 |
|---|---|---|---|
| Bundling 1 |  |  |  |
| m | -1,505 | -1,712 | -1,496 |
| R2 | 0,963 | 0,990 | 0,974 |
| Distr.F | 5,7299E-12 | 3,03587E-09 | 4,48931E-10 |
| Bundling 2 |  |  |  |
| m | -1,027 | -2,138 | -1,792 |
| R2 | 0,974 | 0,964 | 0,947 |
| Distr.F | 5,5066E-13 | 4,67402E-07 | 2,26019E-08 |
| Bundling 3 |  |  |  |
| m | -1,368 | -3,131 | -2,274 |
| R2 | 0,957 | 0,914 | 0,832 |
| Distr.F | 1,8003E-11 | 1,51914E-05 | 1,38949E-05 |
| Bundling 4 |  |  |  |
| m | -1,216 | -2,184 | -1,954 |
| R2 | 0,976 | 0,924 | 0,944 |
| Distr.F | 3,2138E-13 | 9,51187E-06 | 3,19309E-08 |
| Bundling 5 |  |  |  |
| m | -0,212 | -2,340 | -1,734 |
| R2 | 0,041 | 0,922 | 0,853 |
| Distr.F | 0,58986187 | 1,02072E-05 | 6,6603E-06 |

After the convergence period, two new significant events occurred in the Spanish market: operators consolidation, which will be discussed in more detail in section 4.1, and price hike in the main convergent bundles.

Due to the price hikes that began in the 2015 first quarter, mainly in convergent packages and by almost all operators, there is a slowdown in ARPU declines during 2015. In order to isolate this fact, a new analysis period was introduced from 4Q2012 to 1Q2015. This new comparison supports the hypothesis again: there is a higher level of slope decrease (ARPU levels) and higher correlation rates.

Bundle 5, which is composed by FTTH broadband, deserves a special analysis. It presents very low correlation levels for the pre-convergence stage (R2 = 0.04) and barely exceeds the F test for a confidence value α = 0.05, so the hypothesis cannot be validated in this case. In Figure 1, the values of equivalent ARPU for this bundle undergo ups and downs during the pre-convergence period. This may be due to the fact that in that period the deployment of FTTH began in the Spanish market. There were only 18,000 customers with this technology in 2009 and increased up to 60,000 customers in 2010 and 180,000 in 2011. The strong momentum for FTTH deployments (as will be seen in the following hypothesis) began in 2012, along with a stabilization of FTTH ARPUs. However, in the convergence stage, it presents a high correlation with a downward trend of ARPU and passes the F test, but it cannot be compared with the initial period, so the hypothesis for this bundle cannot be validated.

The Spanish Telco market has a deflationary trend compared to the rest of the shopping basket, which is why this fall in prices could be influenced by the evolution of Consumer Price Index (CPI). For this reason, we have compared the evolution of CPI and ARPU per Bundle since 2009 in order to analyze whether this implies a correlation between these two facts or the fall of ARPUs is mainly due to the bundling.

TABLE 6
BUNDLEs ARPUs EVOLUTION vs CPI

|  | 1Q09-3Q12 | 4Q12-1Q15 | 4Q12-4Q15 |
|---|---|---|---|
| Bundling 1 | | | |
| m | -0,377 | -2,094 | -1,874 |
| R2 | 0,973 | 0,981 | 0,975 |
| Distr.F | 1,5754E-13 | 3,53108E-08 | 4,01761E-10 |
| Bundling 2 | | | |
| m | -0,493 | -2,693 | -2,297 |
| R2 | 0,978 | 0,928 | 0,931 |
| Distr.F | 4,3085E-14 | 7,47656E-06 | 9,65304E-08 |
| Bundling 3 | | | |
| m | -0,485 | -3,032 | -2,228 |
| R2 | 0,975 | 0,889 | 0,822 |
| Distr.F | 1,0054E-13 | 4,28526E-05 | 1,93249E-05 |
| Bundling 4 | | | |
| m | -0,517 | -2,148 | -1,969 |
| R2 | 0,980 | 0,866 | 0,916 |
| Distr.F | 2,4554E-14 | 9,28079E-05 | 2,93925E-07 |
| Bundling 5 | | | |
| m | -0,349 | -2,784 | -2,090 |
| R2 | 0,311 | 0,885 | 0,835 |
| Distr.F | 0,00919976 | 5,06864E-05 | 1,27567E-05 |

As was seen in the previous analysis, all bundles have high correlation levels for all periods, except FTTH based bundle, during the pre-convergence period, passing the F-test with a significance level α = 0, 05.

The negative sign of the slopes obtained by OLS indicates that the evolution of the bundle ARPU vs the evolution of the CPI is clearly deflationary. To reinforce this argument, using Jan-2009 as 100%-base, the CPI experienced a 10% cumulative growth in contrast to a 50% drop in Bundle 1 ARPU. If the sign and absolute value of the slope in the pre-convergence stage are analyzed, all bundles show this trend. Observing the convergence period, prior to the prices hike (column 4Q12-1Q15), it can be seen that the absolute values of the slopes are x4 and x8 times higher versus the pre-convergence stage, which clearly indicates that the Spanish Telco market had a much more pronounced fall in the prices than the CPI. Thus, it can be stated that convergent bundling caused a reduction of effective prices much more evenly than the evolution of the other expenses that are included in the families' shopping baskets. Although price hikes (column 4Q12-4Q15), are included the same conclusion can be drawn: the slopes fall between x4 and x6 times compared with the evolution of the CPI.

Finally, the Spanish sector has started with a market repair, although it still has a wide distance to align with the trend of price growth that the shopping basket presents in a regular way.

*Hypothesis 2: Bundling is one of the main stimulus factors for growth and diffusion of broadband.*

Several studies have analyzed the variables that influence the diffusion of broadband, since it is widely recognized as a critical factor for the economic development of a country, as an enabler to increase productivity, growth, innovation, competitiveness and employment of a country (ITU, 2011).

These studies were based on supply and demand models or on innovation diffusion in order to analyze broadband adoption in the Organization for Economic Co-operation and Development (OECD), European Union (EU) or International Telecommunication Union (ITU) member countries, using statistical methods such as cross-sectional Ordinary Least Square (OLS) regression, factor analysis and panel regression. The main identified factors that may affect the adoption of broadband include prior adoption levels, market competitiveness (as measured by Herfindahl-Hirschman Index, HHI, or variations thereof), broadband price, population density, educational levels, Gross Domestic Product (GDP) evolution and the unbundled local loop (ULL) price. In most cases, are cross-country studies (Lin & Wu, 2013), (Dauvin & Grzybowski, 2014), (Turk & Trkman, 2011) and some specific case about broadband diffusion in USA (Deni & Gruber, 2006).

In addition to the linear model, the Bass, Gumpertz and logistic models are typically used to describe the diffusion curve of innovation (S-shape). In Lee & Lee (2010) and Lin & Wu (2013) the Gompertz model is used for the OECD countries in their panel data analysis, including the Arellano-Bond GMM dynamic panel estimation to re-examine the results of previous studies. Turk and Trkman (2011) use the Bass diffusion model to estimate the total number of potential Broadband adopters and the coefficient of innovation associated.

The analysis described in the present study uses a multiple linear regression model together with a logistic diffusion model, in order to identify the main variables influencing the diffusion and growth of Broadband during the convergence period in the Spanish market.

Table 7 shows the variables considered in the models

TABLE 7
VARIABLES AND MEASUREMENTS FOR THE ANALYSIS

| Variable | Measurement |
|---|---|
| Broadband customers | Spanish Market Fixed broadband subscribers |
| Broadband penetration | % fixed broadband subscribers per 100 household |
| FTTH | # FTTH subscribers |
| HFC | # HFC subscribers |
| Bundle | # convergent bundles |
| ARPU | Fixed Broadband ARPU |
| GDP | GDP Quarterly variation |
| HHI BAF | HHI (Herfindahl-Hirschman Idex) for Fixed Broadband |

Table 8 shows the multiple linear regression results (extended analysis), together with the analysis and results (adjusted analysis) by eliminating the variables that have a significance level >5%. It can be clearly seen that the number of convergent bundles, FTTH subscribers and GDP evolution during these quarters are the key variables that influence the Fixed Broadband market growth.

These results corroborate the initial hypothesis, adding relevance to FTTH customer base, which experienced a strong boom during the convergence period, together with a favorable period of economic growth as drivers of Fixed broadband growth.

The logistic diffusion model (Dauvin & Grzybowski, 2014) can be specified as follows: let y stand for the number of broadband customers at time t, expressed as service penetration (broadband penetration over households, in our case) and y* the maximum broadband penetration level for the Market (100% considered). The diffusion curve is given by the following formula:

$$y = \frac{y*}{1+e^{-(a+bt)}} \quad (1)$$

where a is called the location or timing parameter which shifts the diffusion curve backward or forward depending on its sign but without changing the S-shape. The speed parameter, b, measures the diffusion pace and represents the growth rate in number of adopters relative to the number of agents who have not adopted the technology yet.

The following linear equation is obtained after transforming the previous formula and can be analyzed with OLS and multiple linear regression:

$$Ln\left(\frac{y*-y}{y}\right) = -a - bt \quad (2)$$

Table 9 shows the results of logistic diffusion model and presents the same conclusions as previous analyses. The same three variables stand out: evolution of convergent bundles, the number of FTTH subscribers and GDP evolution as fixed broadband diffusion drivers.

The other variables included in the analysis (Extended Analysis), evolution of cable technology (HFC) subscribers, ARPU evolution and Herfindahl-Hirschman Index for fixed broadband present a significance level > 5%.

TABLE 8
RESULTS OF BROADBAND MULTIPLE LINEAR DIFFUSION MODEL

| | Extended Analysis | | | | Adjusted Analysis | | |
|---|---|---|---|---|---|---|---|
| | df | F | Significance F | | df | F | Significance F |
| Regression | 6 | 316,3039675 | 3,70178E-08 | Regression | 3 | 852,861265 | 2,41724E-12 |
| Residual | 7 | | | Residual | 10 | | |
| Total | 13 | | | Total | 13 | | |
| | Coeficient | Standard Error | t Stat | | Coeficient | Standard Error | t Stat |
| Intercept | 11761455,772 | 1843029,551 | 6,382 | Intercept | 11436905,933 | 149055,428 | 76,729 |
| Bundle | 0,102 | 0,030 | 3,428 | Bundle | 0,100 | 0,023 | 4,279 |
| FTTH | 0,282 | 0,234 | 1,204 | FTTH | 0,210 | 0,039 | 5,460 |
| HFC | -0,237 | 1,266 | -0,188 | GDP | 90099,087 | 34295,218 | 2,627 |
| ARPU | 22928,893 | 92471,318 | 0,248 | | | | |
| HHI Fixed BB | -136,883 | 213,206 | -0,642 | | | | |
| GDP | 81390,284 | 42725,593 | 1,905 | | | | |
| R^2 | 0,99633 | | | R^2 | 0,99611 | | |

TABLE 9
RESULTS OF BROADBAND LOGISTIC DIFFUSION MODEL

| | Extended Analysis | | | | Adjusted Analysis | | |
|---|---|---|---|---|---|---|---|
| | df | F | Significance F | | df | F | Significance F |
| Regression | 6 | 310,45 | 3,95E-08 | Regression | 3 | 840,95 | 2,59E-12 |
| Residual | 7 | | | Residual | 10 | | |
| Total | 13 | | | Total | 13 | | |
| | Coeficient | Standard Error | t Stat | | Coeficient | Standard Error | t Stat |
| Intercept | -0,523 | 0,432 | -1,212 | Intercept | -0,529 | 0,035 | -15,190 |
| Bundle | -2,21E-08 | 6,96E-09 | -3,175 | Bundle | -2,12E-08 | 5,47E-09 | -3,874 |
| FTTH | -7,58E-08 | 5,48E-08 | -1,385 | FTTH | -6,43E-08 | 9,00E-09 | -7,143 |
| HFC | -0,017 | 0,010 | -1,661 | GDP | -0,018 | 0,008 | -2,270 |
| ARPU | 2,85E-05 | 4,99E-05 | 0,572 | | | | |
| HHI Fixed BB | -0,007 | 0,022 | -0,326 | | | | |
| GDP | 3,83E-08 | 2,96E-07 | 0,129 | | | | |
| R^2 | 0,9963 | | | R^2 | 0,9961 | | |

4*Hypothesis 3: Bundling has increased the penetration of pay TV*

Traditionally, Spanish pay TV market has been far below other comparable European markets. Table 9 (Offcom, 2007; ANACOM, 2016) shows Spanish market presents 20 p.p. of lower pay TV penetration versus other developed markets like UK, France and Sweden, and is situated only above the Italian market.

TABLE 9
PAY TV PENETRATION (2006)

|  | % Pay-TV penetration |
|---|---|
| Spain | 25% |
| UK | 46% |
| France | 46% |
| Italy | 19% |
| Portugal | 31% |
| Germany | 63% |
| Sweden | 46% |
| USA | 95% |

The different reasons for this low penetration in this market would be an interesting line of research for future investigations. However, as will be analyzed later, the pay TV subscribers have remained stable at roughly 4 Million for more than 6 years (2006 -2012) until the introduction of pay TV within the convergent communications bundles.

Initially, pay TV introduction into communications bundles was done in an opt-in commercial mode, leaving the customer free choice of what type of pay TV packages could be included along with the communications services (mainly films & series, Football, Formula 1, Grand Prix motorcycle racing and others sports). Later, in 2012, hardbundles appear where pay TV is included in the main communications bundles, being configured as reference offer in the Spanish market. That Quad-play offer configured as the usual telecoms proposal to the customers has caused an increase in the number of accesses of pay TV, until reaching the 5.5 Million pay TV subscribers by the end of 2015.

In this section, the main variables that have influenced in pay TV growth are analyzed using multiple linear regression and OLS. In addition, the pay TV subscribers' dependence is also analyzed with the following variables:
- Evolution of pay TV ARPU
- HHI for pay TV market
- Evolution of communications bundles including pay TV services

during two periods, pre-convergence stage (3Q2008-3Q2012) and convergence stage (4Q2012-4Q2015). Table 10 displays the results.

During the pre-convergence period, the variable that influences the growth of pay TV most is the increase in competition (reflected by the HHI evolution), and although the service effective price (ARPU) is reduced by almost 27%, this variable does not present statistical significance to the t-test for a value of 5%. Nevertheless, during the convergence period, where the commercial practice of opt-in and hard-bundle is generalized in the market, the only variable that presents statistical significance for a value of 5% is the number of subscribers with convergent bundles.

TABLE 10
PAY TV GROWTH ANALYSIS

| Pre-convergence Period | | | | Convergence Period | | | |
|---|---|---|---|---|---|---|---|
|  | df | F | Significance F |  | df | F | Significance F |
| Regression | 2 | 46,879 | 2,134E-06 | Regression | 3 | 33,712 | 3,185E-05 |
| Residual | 12 |  |  | Residual | 9 |  |  |
| Total | 14 |  |  | Total | 12 |  |  |
|  | Coeficient | Standard Error | t - stat |  | Coeficient | Standard Error | t-stat |
|  |  |  |  | Bundle TV | 0,590 | 0,153 | 3,855* |
| HHI | -594,766 | 81,910 | -7,261* | HHI | -15,572 | 103,437 | -0,151 |
| ARPU TV | 10816,333 | 7890,813 | 1,371 | ARPU TV | 1066,760 | 52095,788 | 0,020 |
| R^2 |  | 0,887 |  | R^2 |  | 0,9183 |  |

*Significance at 5% level

(3) Source: Comisión Nacional de los Mercados y de la Competencia

**4.2 Market structure**

*Hypothesis 4   Bundling has encouraged consolidation in the Spanish Telco Market*

The introduction and subsequent widespread growth of telecommunications bundling, FTTH´s deployments and its inclusion in the connectivity offer, together with pay TV and communications bundling by the main players, caused that these offers were placed as reference offer or value proposition at the Spanish market.

This pay TV and communications bundling as "facto" reference offer led the main players to need these capabilities, either directly, through their own assets and networks, complemented with a wholesale offer, or through other companies' acquisitions.

The launch of convergent bundles by the main market player (Telefónica), as well as its commitment to FTTH deployments, placed it in a clear competitive advantage versus its main competitors (Vodafone and Orange).

These capabilities needed to compete motivated other market agents acquisition seeking access to Ultra-Broadband immediate capabilities, deployments already initiated in FTTH or acquire capacities, positioning and know-how in the pay TV market. Therefore, the following acquisitions[4] were made in the market:

- Vodafone acquired ONO (Jul-2014) providing access to an HFC footprint of 8 Million Households.
- Orange acquired Jazztel, (Jun-2014), giving access to 2.2 Million Households with FTTH.
- Telefonica acquired Canal +, (April-2015) providing access to the largest pay TV customer base and DTH capabilities.
- Recently, Grupo Masmovil acquired Yoigo, (Sept-2016) beginning to deploy FTTH, with the aim of becoming a convergent operator.

The way these acquisitions have impacted on the market structure can be analyzed through the evolution of HHIs in the different stand-alone markets: Telephony, Internet Services, Mobile communications, and pay TV.

.

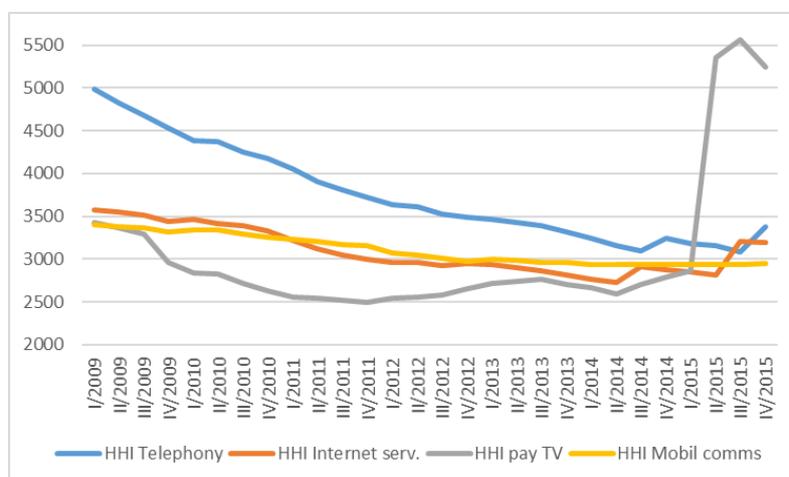

Fig. 2. Herfindahl-Hirschman indices evolution for Telephony, Internet Services, pay TV and Mobile communications

Figure 2 shows that these indices present a downward path due to the increase market competition in all services. This tendency has been slowed or increased dramatically (e.g., pay TV), due to the effects of agents' consolidation discussed above.

But since the element analyzed is the convergent market, Herfindahl-Hirschman Index Product-to-Product does not show the whole impact of bundling and therefore a ratio that shows how the market has been concentrated around the main convergent players was specifically designed for this purpose by the authors of this paper.

The concentration ratio has been defined as the number of accesses fixed telephony, broadband, pay TV and mobile communications access of the three main convergent players -Telefonica, Vodafone

(4)  Source: Companies reports

and Orange- (Acc_ConvOpe) over total accesses of these products market (FT, BB, MA, pay TV.

$$\%Concentration = \frac{\sum_i Acc\_ConvOp_i}{\sum_{j,k,l,m} FT_j + BB_k + MA_l + PayTV_m} \quad (3)$$

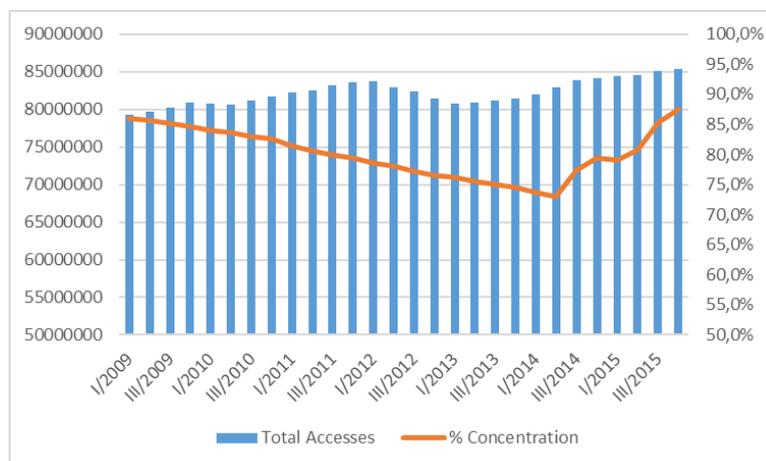

Fig. 3. Evolution of total accesses in the Spanish market and %concentration in main convergent agents

Figure 3 shows how, since the beginning of the series in 1Q2009, the concentration ratio has experienced a clear downward trend, going from 86% (1Q2009) to 73% (2Q2014), an inflection point where it rises to 87.4% in 4Q2015, with a value even higher than the one obtained in the beginning of 2009.

## 5  CONCLUSIONS

This article aimed to identify the main impacts of bundling and convergence in the Spanish Telco Market. Introduction and subsequent massification of convergent bundling have increased the level of price discount, causing an incremental fall in the effective price of communications products that make up the market. Additionally, it was confirmed that this level of discount is higher than the tendencies already experienced by the consumer price index. Another benefit corroborated by the article is the greater dynamisation of the broadband market, causing higher diffusion levels in opposition to previous growth models. Moreover, the introduction of payTV within bundling has led to an unprecedented increase of penetration for this product.

Finally, it was also demonstrated how the need to have the necessary network capabilities to market convergent bundles and to ensure market competitiveness has led to a concentration of the sector around large convergent operators.